%Paper: hep-th/9301104
%From: harnad@MATHCN.UMontreal.CA (Harnad John)
%Date: Mon, 25 Jan 1993 17:40:57 -0500 (EST)
%Date (revised): Thu, 28 Jan 1993 00:12:34 -0500 (EST)

%%%%%%%%%%%%%%%%%%%%%%%%%%%%%%%%%%%%%%%%%%%%%%%%%%%%%%%%%%%%%%%%%%%%%%%%%%
% Title:  Moment Maps to Loop Algebras, Classical R-Matrix
%         and Integrable Systems
% Author: J. Harnad and M.-A. Wisse
% Date:  Jan. 23, 1993
%=============================================================
% Type:  Conference Proceedings
%=============================================================
% NSERC-CAP Summer Institute in Theoretical Physics
% Workshop on Quantum Groups, Integrable Models and Statistical Systems
% Kingston Canada, July 13 - 17th 1992
%=============================================================
% Compiler:  AMSTeX version 2.1 or later
% ===========================================================
%%%%%%%%%%%%%%%%%%%%%%%  Formatting Specifications  %%%%%%%%%%%%%%%%%%%%%
\documentstyle{amsppt}
%\NoPageNumbers
\TagsOnRight
%\nologo
% The following 3 lines "disable" the printing of the AMSTeX logo.
% Please be sure that proper acknowledgment is made upon publication.
\catcode`\@=11
\def\logo@{}
\catcode`\@=13
\parindent=8 mm
\magnification 1200
\hsize = 6.75 true in
\vsize = 8.5 true in
%\hoffset = .2 true in
\parskip=\medskipamount
\baselineskip=14pt
%%%%%%%%%%%%%%%%%%%%%%%%%%%% Definitions %%%%%%%%%%%%%%%%%%%%%%%%%%%%%%%%
\def \Ad{\rom{Ad}}
\def\rom{\roman}

\def \smaller {\eightpoint}
\def \wt {\widetilde}
\def \wh {\widehat}
\def \mt {\mapsto}
\def \ra {\rightarrow}

\def \lra {\longrightarrow}
\def \lmt {\longmapsto}
\def \a {\alpha}
\def \b {\beta}

\def \e {\epsilon}
\def \g {\gamma}
\def \G {\Gamma}

\def \l {\lambda}

\def \ph {\varphi}

\def \s {\sigma}
\def \S {\Sigma}

\def \t {\tau}
\def \o {\omega}

\def \ss {\subset}
\def \Lg {\widetilde{\frak g}}
\def \LG {\widetilde{\frak G}}
\def \LGp {\widetilde{\frak G}^+}
\def \LGm {\widetilde{\frak G}^-}
\def \LGpm {\widetilde{\frak G}^{\pm}}
\def \Lgp {\widetilde{\frak g}^+}
\def \Lgm {\widetilde{\frak g}^-}
\def \Lgpm {\widetilde{\frak g}^{\pm}}
\def \lgp {\widetilde{\frak g}_+}
\def \lgm {\widetilde{\frak g}_-}

\def \lgpm {\widetilde{\frak g}_{\pm}}
\def \gYA {{\frak g}_{A}^Y}
\def \IYA {{\Cal I}_A^Y}

\def \CC {\Cal C}

\def \JJ {\Cal J}
\def \KK {\Cal K}
\def \LL {\Cal L}

\def \NN {\Cal N}
\def \LL {\Cal L}

\def \ln{\text{ln}}
\def \tr{\text{tr }}
\def \diag{\text{diag }}
\def \atil{\wt{a}}
\def \btil{\wt{b}}
\def \ctil{\wt{c}}
\def \ztil{\wt{z}}
\def \Ptil{\wt{P}}
\def \Ctil{\wt{\CC}}

%%%%%%%%%%%%%%%%%%%%%%%%% Title, Author & Abstract  %%%%%%%%%%%%%%%%%%%%%%%%%%
\phantom{0}\vskip -.5 true in \rightline{CRM-1854 (1993)  }
\vskip .5 true in
\topmatter
\title
 Moment Maps to Loop Algebras \\ Classical $R$--Matrix
and Integrable Systems${}^\dag$
\endtitle
\rightheadtext{Moment Maps to Loop Algebras}
\leftheadtext{J. Harnad and M.-A.~Wisse}
\endtopmatter
\footnote"" {\hskip -.4 true in${}^\dag$ Talk given by J.~H.
 at the NSERC-CAP Summer Institute in Theoretical Physics,
 Workshop on Quantum Groups, Integrable Models and Statistical Systems.
 Kingston, Canada, July 13 - 17th 1992.
 Research supported in part by the Natural Sciences and Engineering Research
 Council of Canada and the Fonds FCAR du Qu\'ebec.}
\footnote""{\hskip -.4 true in$^1$e-mail address:
harnad\@alcor.concordia.ca  {\it \ or\ }  harnad\@mathcn.umontreal.ca}
\footnote""{\hskip -.4 true in$^2$e-mail address:
wisse\@mathcn.umontreal.ca}
%%%%%%%%%%%%%%%%%%%%%%%%%%%%%%%%%%%%%%%%%%%%%%%%%%%
\centerline{\smc J.~Harnad${}^1$}
\smallskip
{\smaller \centerline{ \it Department of Mathematics and Statistics, Concordia
University }
\centerline{ \it  7141 Sherbrooke W., Montr\'eal, Canada H4B 1R6, \
and }
\centerline{ \it  Centre de recherches math\'ematiques,
Universit\'e de Montr\'eal }
\centerline{ \it  C.~P. 6128-A, Montr\'eal, Canada
H3C 3J7}}
\medskip
{\smaller\centerline{and}}
\medskip
 \centerline{\smc M.~A.~Wisse${}^2$}
\smallskip
{\smaller \centerline{ \it D\'epartement de math\'ematiques
et de statistique,  Universit\'e de Montr\'eal}
\centerline {\it C.P. 6128-A, Montr\'eal, P.Q., Canada H3C 3J7 }}
\bigskip \bigskip
%%%%%%%%%%%%%%%%%%%%%%%%%%%%%%%%%%%%%%%%%%%%%%%%%%
\centerline{\bf Abstract}
\bigskip
\baselineskip=10pt
\centerline{
\vbox{
\hsize= 5.5 truein
{\smaller
A class of Poisson embeddings of reduced, finite dimensional
symplectic vector spaces into the dual space $\Lg_R^*$ of a loop algebra, with
Lie Poisson structure determined by the classical split $R$--matrix $R=P_+ -
P_-$ is introduced. These may be viewed as equivariant moment maps inducing
natural Hamiltonian actions of the ``dual'' group $\LG_R = \LGp \times \LGm$ of
a loop group $\LG$ on  the symplectic space. The $R$--matrix
version of the Adler-Kostant-Symes theorem is used to induce commuting flows
determined by isospectral equations of Lax type. The compatibility conditions
determine finite dimensional classes of solutions to integrable systems of
PDE's, which can be integrated using the  standard Liouville-Arnold
approach. This involves an appropriately chosen ``spectral Darboux''
(canonical)
coordinate system in which there is a complete separation of variables. As an
example, the method is applied to the determination of finite dimensional
quasi-periodic solutions of the sine-Gordon equation.  }}}
 \baselineskip=14pt
\bigskip
%%%%%%%%%%%%%%%%%%%%%%%%%%%%%%% Section 0  %%%%%%%%%%%%%%%%%%%%%%%%%%%%%%%%
\subheading{0. Introduction}

   In {\bf[AHP, AHH1-AHH3]} a unified framework was developed, describing a
wide class of integrable finite dimensional Hamiltonian systems, as well as
finite dimensional solutions to integrable systems of PDE's. The basic idea is
to represent all these systems in terms of commuting isospectral flows
determined by Lax equations within finite dimensional Poisson subspaces of loop
algebras consisting of orbits that are rational in the loop parameter. The
commuting flows are generated by Hamiltonians that are spectral invariants, or
equivalently $Ad^*$--invariants on the space $\Lg^{+*}$  dual to the half of
the loop algebra admitting holomorphic extensions to the interior disk of a
circle within the complex plane of the loop parameter. According to the
Adler-Kostant-Symes (AKS) theorem, such invariants generate commuting flows
with
respect to the Lie Poisson structure on $\Lg^{+*}$ and are determined by
equations of Lax type.

   Finite dimensional, generically quasi-periodic solutions to integrable
systems of PDE's arise within this framework as the compatibility conditions
satisfied by the matrix elements, within a given representation. The procedure
developed in {\bf[AHP, AHH1-AHH3]} consists of three steps; first, a
suitably defined moment mapping is used to embed symplectic vector spaces
quotiented by Hamiltonian group actions as Poisson subspaces of the space
$\Lg^{+*}$. The image consists of rational coadjoint orbits with fixed poles.
The AKS theorem is then used to define the Hamiltonians and determine the
dynamical equations, with the ring of invariants generated by the invariant
spectral curve given by the characteristic equation of the loop algebra
element. The generic systems so obtained may be subjected to further reductions
under both discrete and continuous Hamiltonian symmetry groups. Finally, a
suitably defined Darboux (canonical) coordinate system is introduced,
associated to the divisor of zeroes of the matrix representating the loop
algebra element on the spectral curve. Within this ``spectral Darboux''
coordinate system, there is a complete separation of variables which, through
application of the Liouville-Arnold integration method on the invariant
Lagrangian manifolds, reduces the problem of integration to quadratures given
by Abelian integrals. The resulting flows are thus seen to be linearized by the
Abel map in the  natural linear coordinate system on the Jacobian variety of
the spectral curve, or some quotient thereof. Through the Jacobi inversion
method, the matrix elements may be expressed in terms of theta functions,
thereby producing, through classical Hamiltonian methods, the types of
solutions
typically  obtained through more sophisticated  algebro-geometric means
{\bf [AvM, KN, D, AHH1] }.

	This approach has been applied to a wide variety of integrable systems, both
finite and infinite dimensional, such as: the cubically nonlinear Schr\"odinger
equation and various multi-component generalizations thereof {\bf [AHP,
AHH3, HW1, W1]}, the sine-Gordon equation {\bf [HW2, TW]}, the massive
Thirring model {\bf [W2]} the Neumann and Rosochatius systems {\bf [AHP,
AHH1, AHH3]} and various generalized tops {\bf [AHH2, HH]}.

   In the subsequent section, an extended form of the moment map embedding
method of {\bf [AHP, AHH2]} will be given, based on the classical $R$--matrix
technique, in which the two halves of the loop algebra $\Lgp$, $\Lgm$ are
treated  on an equal footing. In section 3, the process of discrete reduction,
together with the Liouville-Arnold linearization method, will be illustrated
for the case of the sine-Gordon equation. Full details for the latter may be
found in {\bf [HW2]}.
\bigskip
 %%%%%%%%%%%%%%%%%%%%%%%%%%%%%%% Section 1 %%%%%%%%%%%%%%%%%%%%%%%%%%%%%%%%

\subheading{1. Moment Maps to Loop Algebras and the Classical $R$--Matrix }

We first establish notations for loop groups and algebras. Let
$(M,\o)$ denote the  symplectic vector space whose elements are pairs  $(F,G)$
of $N \times r$ matrices
$$
M=\{(F,G)\in M^{N,r} \times M^{N,r}\}, \tag{1.1}
$$
with symplectic form:
$$
\o= \tr (dF^T\wedge dG).  \tag{1.2}
$$
Let $\LG$ denote the group of smooth loops in $\frak{Gl}(r)$ (real or
complex), or some subgroup thereof, viewed as invertible matrix-valued smooth
functions $g(\lambda)$ defined on a smooth, simple closed curve $\G$ enclosing
the origin of the complex $\lambda$  plane.  Let $\LGp$, $\LGm$ be the
subgroups of loops admitting  holomorphic extensions, respectively, to the
interior and exterior regions $\G^{+}, \ \G^{-}$  (including $\infty$), such
that for $g\in \LGm$, $g(\infty)=I$. The  Lie algebras corresponding to $\LG$,
$\LGpm$, are denoted $\Lg$ and $\Lgpm$ respectively. Their elements are smooth
$\frak{gl}(r)$ valued functions on $\G$, with elements $X_+\in \Lgp$ admitting
holomorphic extensions to $\G^+$ and $X_-\in \Lgm$ to $\Gamma^-$, the latter
satisfying $X(\infty) =0$.

     We identify $\Lg$ as a dense subspace of the dual space
$\Lg^*$ through the  pairing:
 $$
 \align <X, Y> &:=
\frac{1}{2\pi i} \oint_{\G}\tr\left(X(\l)Y(\l)\right)\frac{d\l}{\l} ,
\\ X \in \Lg^* &\ , \ Y\in \Lg. \tag{1.3}
\endalign
$$
Using the vector space decomposition
$$
\Lg = \Lgm \oplus \Lgp, \tag{1.4}
$$
this allows us to identify the dual spaces as
$$
\Lg^{+*} \sim \lgm, \qquad \Lg^{-*} \sim \lgp,  \tag{1.5}
$$
where $\lgm$, $\lgp$ denote, respectively, the subspaces of $\Lg$ consisting of
elements admitting a holomorphic extension to $\G^-$ or  $\G^+$,
with $X_+ \in \lgp$ satisfying $X_+(0)=0$.

 Let
$$
\aligned
P_{\pm}:\Lg &\lra \Lgpm \\
 P_{\pm}: X & \lmt X_{\pm}
\endaligned      \tag{1.6}
$$
be the projections to the subspaces $\Lgpm$ relative to the decomposition (1.4)
and define the endomorphism $R:\Lg \ra \Lg$ as the difference:
$$
R := P_+ - P_-.      \tag{1.7}
$$
Then $R$ is a {\it classical R-matrix} {\bf[S]}, in the sense that the bracket
$[\ ,\ ]_R$ defined on $\Lg$ by:
$$
[X,Y]_R:= {1\over2}[RX,Y] + {1\over2}[X,RY]       \tag{1.8}
$$
is skew symmetric and satisfies the Jacobi identity, determining a
new Lie algebra structure on the same space as $\Lg$, which we denote $\Lg_R$.
The corresponding group is the ``dual group'' $\LG_R = \LGm \times \LGp$
associated to $\LG$.  The adjoint and coadjoint actions of $\LG_R$ on
 $\Lg_R \sim \Lg_R^*$ are given by:
$$
\align
\Ad_R(g): (X_ + +X_-) &\lmt g_+X_+g_+^{-1} + g_-X_-g_-^{-1}  \tag{1.9a}\\
\Ad^*_R(g): (Y_+ +Y_-) &\lmt (g_-X_+g_-^{-1})_+ + (g_+X_-g_+^{-1})_-,
\tag{1.9b}\\
X_{\pm} \in \Lgpm, & \qquad Y_{\pm} \in \lgpm, \qquad   g_{\pm} \in \LGpm.
\tag{1.9}     \endalign
$$
The Lie Poisson bracket on $\Lg_R^* \sim \Lg^*$ dual to the Lie bracket
$[\ ,\ ]_R$ is:
$$
\{f, g\}\vert_X =<[df, dg]_R,X>       \tag{1.10}
$$
for smooth functions $f,g$ on $\Lg_R^*$ (with the usual identifications,
$df\vert_X,\ dg\vert_X \in \Lg^* \sim \Lg)$.

    Let $A \in \frak{gl}(N)$ be a fixed $N \times N$ matrix having no
eigenvalues  on $\G$, and define the following action of $\LG_R$ on $M$
$$
\align
\LG_R: M &\lra M \\
g(\l):(F,G) &\lra (F_g, G_g) \tag{1.11}
\endalign
$$
where $(F_g, G_g)$ are defined by:
$$
\align
F_g &:= F - {1\over 2\pi i} \oint_{\G} (A-\l I)^{-1} F
\left(g_+^{-1}(\l) - g_-^{-1}(\l) \right)d\l  \tag{1.12a}
\\
G_g &:= G - {1\over 2\pi i} \oint_{\G} (A^T-\l I)^{-1} G
\left(g_+^{T}(\l) - g_-^{T}(\l)\right)d\l.  \tag{1.12b}
\endalign
$$
It is straightforward to verify, using the identity
$$
(A-\l I)^{-1}(A-\s I)^{-1} = {(A-\l I)^{-1} - (A-\s I)^{-1} \over\l-\s}
\tag{1.13}
$$
and residue calculus, that the $\LG_R$ composition rule is  satisfied, so
(1.11), (1.12a,b) does, indeed define a $\LG_R$--action on $M$. A similar
calculation shows that this action preserves  the symplectic form (1.2) and
is, in fact, generated as a Hamiltonian action by the equivariant moment map:
$$
\align
\wt{J}_{A,Y} : M &\lra \Lg_R^* \\
 \wt{J}_{A,Y}  (F,G) &=  \l Y + \l G^T(A- \l I)^{-1}F , \tag{1.14}
\endalign
$$
where $Y\in\frak{gl}(r)$ is any $r\times r$ matrix. The constant term
$\l Y$ may be included without affecting the equivariance of the moment map
 (1.14) since $\l Y$ is an infinitesimal character for the Lie algebra $\Lg_R$:
$$
<\l Y, [X,Y]_R> = 0, \quad \forall X, Y \in \Lg_R.   \tag{1.15}
$$
The splitting $\wt{J}_{A,Y}(F,G) = \wt{J}_{+}(F,G) + \wt{J}_{-}(F,G)$,
$\wt{J}_{\pm}(F,G) \in \lgpm$ is determined by the pole structure, with the
$\l Y$ term plus the poles at eigenvalues of $A$ in $\G^-$ included in the
$(\Lgm)^*\sim\lgp$ part and the poles at eigenvalues in $\G^+$ in the
$(\Lgp)^*\sim\lgm$ part.

The fibres of the Poisson map $\wt{J}_{A,Y}$ are the orbits of the stability
subgroup $G_A := \text{Stab}(A) \ss Gl(N)$ under the Hamiltonian $Gl(N)$
action defined  by
$$
\align
g: M &\lra M \\
g:(F,G) &\lmt (gF, (g^T)^{-1}G). \tag{1.16}
\endalign
$$
On a suitably defined open, dense set, this action is free, allowing us to
define the quotient Poisson space $M/G_A$ . Since the map $\wt{J}_{A,Y}:M \ra
\Lg^*_R$ passes to the quotient, defining a $1$--$1$ Poisson map
$\wt{J}_{A,Y}:M/G_A \ra \Lg^*_R$, we may identify $M/G_A$  with the image
Poisson space $\gYA \ss \Lg^*_R$ consisting of elements $ {\Cal N}(\l) $ of the
form
$$
{\Cal N}(\l) =  \l Y +
 \l\sum_{i=1}^n \sum_{l_i=1}^{n_i} {N_{i,l_i} \over(\l -\a_i)^{l_i}},
\tag{1.17}
$$
where $\{\a_i\}_{i=1, \dots n}$ are the eigenvalues of $A$,
 $\{l_i\}_{i=1, \dots n}$ are the dimensions of the corresponding Jordan blocks
of generalized eigenspaces, and the ranks and Jordan structures of the $r\times
r$ matrices $N_{i, l_i}$ are determined by the multiplicities of the
eigenvalues of $A$ and its Jordan structure.

The Hamiltonian equations on the phase space $\gYA \sim M/G_A$ are generated by
elements of the ring $\IYA$ consisting of $Ad^*$--invariant polynomials on
$\Lg^*$, restricted to the Poisson subspace $\gYA \ss \Lg^*_R$. According to
the Adler-Kostant-Symes (AKS) theorem, in its $R$--matrix form {\bf[S]}, the
elements of this ring Poisson commute, generating commuting Hamiltonian flows,
and Hamilton's equations for $\phi \in \IYA$ have the Lax form:
$$
\align
{d\Cal{N}(\l)\over dt} &= [(d\phi)_+, \Cal{N}] = -[(d\phi)_-, \Cal{N}], \\
 (d\phi)_{\pm}\in \Lgpm,& \quad d\phi = (d\phi)_+ + (d\phi)_-.  \tag{1.18}
\endalign
$$
This implies that the spectral curve $\Cal S$, with affine part defined by the
characteristic equation
$$
\det(\Cal{L}(\l) - z I) := \Cal{P}(\l,z) = 0,  \tag{1.19}
$$
where
$$
\Cal{L}(\l) := {a(\l)\over \l} \Cal{N}(\l), \qquad a(\l)
:= \prod_{i=1}^{n}(\l - \a_i)^{n_i}
\tag{1.20}
$$
is invariant under these flows, and its coefficients generate the ring $\IYA$.

 Integrable systems of PDE's then arise as the equations satisfied by the
matrix elements of $\Cal{N}$ implied by the compatibility conditions
$$
\align
{d(d\phi)_+\over dx} - {d(d\psi)_+\over dt}& +[d(\phi)_+,d(\psi)_+] =0,
\tag{1.21} \\
\phi, \ \psi & \in \IYA,
\endalign
$$
where $t, \ x$ are the respective flow parameters for the Hamitonians $\phi$
and $\psi$. The flows may be integrated through a standard algebro-geometric
construction that leads to a linearizing map to the Jacobi variety of the
spectral curve $S$  ({\bf[KN, D, AHH1]}). To obtain interesting
examples, the generic systems so obtained must usually be further reduced by
certain continuous or discrete symmetry groups. The easiest way to
arrive at the linearization involves the introduction of a suitably defined
``spectral Darboux'' (canonical) coordinate system associated to the divisor of
zeroes of the eigenvectors of $\Cal{L}(\l)$ on the spectral curve. The general
construction is detailed in {\bf [AHH3]}. Its application to the particular
problem of determining finite dimensional quasi-periodic solutions to the
sine-Gordon equation is described in the next section. Full details for this
case may be found in  {\bf [HW2]}.
\bigskip
 %%%%%%%%%%%%%%%%%%%%%%%%%%%%%%% Section 2 %%%%%%%%%%%%%%%%%%%%%%%%%%%%%%%%
\subheading{2. Quasiperiodic Solutions of the Sine-Gordon Equation}

  To obtain the sine-Gordon equation
$$
\frac{\partial^2u}{\partial x^2}-\frac{\partial^2u}{\partial t^2}
= \sin(u) \tag{2.1}
$$
we shall need commuting isospectral flows in the twisted loop algebra
$\wh{\frak{su}}(2) :=\wt{\frak{su}}^{(1)}(2)$. This is the subalgebra of
$\wt{\frak{gl}}(2)$ consisting of elements $X(\l)$ that are invariant
under the three involutive endomorphisms:
$$
\align
\s_1: X(\l) &\lmt  J X^{T}({\l})J  \tag{2.2a} \\
\s_2: X(\l) &\lmt - X^{\dag}(\bar{\l})   \tag{2.2b} \\
\s_3: X(\l) &\lmt \t X (-\l)\t  \tag{2.2c}
\endalign
$$
where
$$
\t := \pmatrix 1 & 0 \\ 0 & -1 \endpmatrix, \quad
J := \pmatrix 0 & -1 \\ 1 & \phantom{-}0 \endpmatrix. \tag{2.3}
$$
The first of these implies that $X$ is traceless, the second that
it is in $\wt{\frak{su}}(2)$ and the third that it is in the
``twisted'' subalgebra $\wh{\frak{su}}(2)$.

   We  choose the matrix $Y$ as
$$
Y = J = \pmatrix 0 & -1 \\ 1 & \phantom{-}0 \endpmatrix, \tag{2.4}
$$
$\Gamma$ as a circle centred at the origin of the $\l$--plane, and $A$ as  a
diagonal $2N \times 2N$ dimensional matrix, having distinct eigenvalues, all
inside $\G^+$, of the form
$$
A = \diag (\a_1, \bar{\a}_1, \dots ,\a_p, \bar{\a}_p, -\a_1,
\dots , -\bar{\a}_p, i \b_{2p+1}, -i\b_{2p+1},  \dots, i \b_N,
-i\b_N),   \tag{2.5}
$$
where $\{\a_j\}_{j=1, \dots, p}$ have nonvanishing real and imaginary parts
and   $\{\b_j\}_{j=2p+1, \dots, N}$ are real.

   The space $M=\{(F,G)\in M^{2N,2} \times M^{2N,2}\}$ must  correspondingly be
reduced by the endomorphisms
$$
\align
\S_1:(F,G) &\lmt (GJ, -FJ) \tag{2.6a}\\
\S_2:(F,G) &\lmt (-\JJ \bar G, \JJ \bar F) \tag{2.6b}\\
\S_3:(F,G) &\lmt (-i \KK F\t, i\KK G\t) \tag{2.6c}\\
\endalign
$$
where
$$
\JJ = \diag(J, J, \dots, J)  \tag{2.7}
$$
is the block diagonal $2N \times 2N$ matrix with $2 \times 2$ blocks equal to
$J$ and
$$
\KK = \pmatrix \phantom{-}0 & I_{2p} &    &        &    \\
                    -I_{2p} &    0   &    &        &    \\
                            &        &  J &        &    \\
                            &        &    & \ddots &    \\
                            &        &    &        &  J
\endpmatrix   \tag{2.8}
$$
is the $2N \times 2N$ matrix consisting of a $4p \times 4p$  block formed from
$I_{2p}$, the  $2p \times 2p$ identity matrix and its negative,  and
$N-2p$ diagonal $2\times 2$ blocks $J$.

The moment map $\wt{J}^Y_A$ of eq.~(1.14) then intertwines the automorphism
groups generated by $\S_1, \S_2, \S_3$ and by $\s_1, \s_2, \s_3$. Denoting by
$\left(\pmatrix F_{2j-1} \\ F_{2j}\endpmatrix, \  \pmatrix G_{2j-1}\\
G_{2j}\endpmatrix \right)$ the consecutive  pairs of $2 \times 2$ blocks in
$(F,G)$, the fixed point set $M_\S \ss M$ under
 $\S_1, \S_2, \S_3$ consists of $(F,G)$ with $2\times 2$ blocks of the form
$$
\align
\pmatrix F_{2j-1} \\ F_{2j}\endpmatrix &=
\pmatrix \ph_j &\bar \g_j\\ \g_j & -\bar\ph_j \endpmatrix, \quad
\pmatrix G_{2j-1}\\ G_{2j}\endpmatrix =
\pmatrix -\bar\g_j & \ph_j \\ \bar\ph_j &  \g_j\\  \endpmatrix,
\quad j=1, \dots,p\tag{2.9a}\\
\pmatrix F_{2j+2p-1} \\ F_{2j+2p}\endpmatrix &=
\pmatrix i\ph_j & -i\bar\g_j\\ i\g_j & i\bar\ph_j \endpmatrix, \quad
\pmatrix G_{2j+2p-1} \\ G_{2j+2p}\endpmatrix =
\pmatrix i\bar\g_j & i\ph_j \\ -i\bar\ph_j &  i\g_j\\  \endpmatrix,
\quad j=1, \dots,p \tag{2.9b}\\
\pmatrix F_{2j-1} \\ F_{2j}\endpmatrix &=
\pmatrix i \g_j &\bar\g_j\\ \g_j & i \bar\g_j \endpmatrix, \quad
\pmatrix G_{2j-1} \\ G_{2j}\endpmatrix =
\pmatrix -\bar\g_j & i\g_j  \\ -i \bar\g_j &  \g_j \endpmatrix,
\quad
j= 2p +1, \dots ,N. \tag{2.9c}
\endalign
$$

 We have
$$
\S_1^* \o = \o,\quad \S_2^* \o = \bar{\o}, \quad  \S_3^*{\o} =\o
\tag{2.10}
$$
so $M_\S$ is a real symplectic space, with symplectic form:
$$
\wh \o=
4\sum_{j=1}^p(d\gamma_j\wedge d\bar\varphi_j
+ d\bar\gamma_j \wedge d\varphi_j) +
4i \sum_{j=2p+1}^N d\bar\gamma_j\wedge d\gamma_j. \tag{2.11}
$$
The restriction of $\wt{J}^Y_A$ to $M_\s$, which we denote $\wh{J}$, is given
by
$$
\wh{J}(F,G) :=\NN(\l) = \l\pmatrix 0 & -1 \\ 1 & 0 \endpmatrix +
 2\l \pmatrix  b(\l) & -c(\l) \\ \bar{c}(\bar\l) & -b(\l) \endpmatrix
\in \wh{\frak{su}}(2).
 \tag{2.12}
$$
where
$$
\aligned
b(\l)&=\l\sum_{j=1}^p \left(\frac{-\varphi_j \bar\gamma_j}{\a_j^2-\l^2} +
\frac{\bar\varphi_j \gamma_j}{\bar\alpha_j^2-\l^2}\right) +
i\l\sum_{j=2p+1}^N \frac{\left\vert\gamma_j\right\vert^2}{\b_j^2+\l^2}\\
c(\l) &=\sum_{j=1}^p \left(\frac{\a_j\bar\gamma_i^2}{\a_j^2-\l^2}+
\frac{\bar{\alpha}_j\bar\varphi_j^2}{\bar\alpha_j^2-\l^2}\right)
-i\sum_{j=2p+1}^N\frac{\b_j\bar\gamma_j^2 }{\b_j^2+\l^2}.
\endaligned
\tag{2.13}
$$

  Now, choose Hamiltonians $H_\xi, H_\eta \in \IYA$ as:
$$
\align
H_\xi(X) &=-{1\over 4\pi i}\oint_\G  \tr\left(\frac{a(\l)}{\l^2}(X(\l)
)^2\right) d\l    \tag{2.14a}\\
 H_\eta(X)&={1\over 4\pi i}\oint_\G
\tr\left(\frac{a(\l)}{\l^{2N}}(X(\l))^2 \right) d\l,  \tag{2.14b}
\endalign
 $$
where
$$
a(\l)=\prod_{j=1}^p[(\l^2-\a_j^2)(\l^2-\bar\alpha_j^2)]
\prod_{k=2p+1}^N(\l^2+\b_k^2). \tag{2.15}
$$
Then the Lax form of Hamilton's equations may be written as
$$
\align
{d\NN \over d\xi} &= -\left[dH_\xi(\NN)_-,\NN \right]  \tag{2.16a} \\
{d\NN \over d\eta} &= \left[dH_\eta(\NN)_+,\NN \right],  \tag{2.16b}
\endalign
$$
where, setting
$$
\LL(\l)=\frac{a(\l)}\l \Cal N(\l)=\l^{2N-1}\LL_0+\l^{2N-2}\LL_1+\dots+
\LL_{2N-1}+a(\l)Y,\tag{2.17}
$$
we have
$$
\align
dH_\xi(\Cal N)_-&=-\frac 1\l (\LL_{2N-1} + a(0)Y)   \tag{2.18a}\\
dH_\eta(\Cal N)_+&= \LL_0 + \l Y.  \tag{2.18b}
\endalign
$$
The spectral curve is of the form
$$
\gather
\det(\LL (\l) -z I) = \Cal P(\l,z)=z^2+a(\l)P(\l)=0 \tag{2.19a} \\
P(\l)=P_0+\l^2 P_1 + \dots + \l^{2N-2} P_{N-1} +\l^{2N}, \tag{2.19b}
\endgather
$$
with all coefficients $P_i$ in $\IYA$ and, in particular
$$
H_\xi=P_0,\qquad H_\eta=-P_{N-1}. \tag{2.20}
$$
Choosing the invariant level set
$$
a(0)P_0=\det(\LL_{2N-1}+a(0)Y) ={1\over 16}, \tag{2.21}
$$
implies that
$$
\LL_{2N-1}+a(0)Y= {1\over 4} \pmatrix 0 & e^{i u} \\ -e^{-i u} & 0
\endpmatrix, \tag{2.22}
$$
where $u$ is real. Setting
$$
\xi = x + t, \quad \eta = x - t, \tag{2.23}
$$
the compatibility equations for (2.16a,b) then imply that $u$ satisfies the
sine-Gordon equation (2.1).

Since  the spectral curve $\CC$ determined by (2.19a,b) is invariant under the
involution $(z,\l)\mt(z,-\l)$, it is a two-sheeted covering of the
hyperelliptic curve with genus $N-1$ whose affine  part given is by
$$
z^2+\wt{a}(E)\wt{P}(E)=0, \tag{2.24}
$$
where
$$
\wt{P}(\l^2):=P(\l), \quad \wt{a}(\l^2):=a(\l), \quad \l^2=E. \tag{2.25}
$$
We also define functions $\btil , \ \ctil$ by
$$
\btil(\l^2)=b(\l),\quad \ctil(\l^2)=c(\l). \tag{2.26}
$$
Setting $\ztil:=z\l$,  we obtain the genus $N$ hyperelliptic curve
$\wt{\CC}$ with affine part given by
$$
\ztil^2+E\atil(E)\Ptil(E)=0, \tag{2.27}
$$
which has two additional branch points at $E=0, \infty$.

   Following the general method of {\bf [AHH3]}, we define on $\wt{\CC}$ the
divisor of degree $N$ with coordinates $(E_\mu,\zeta_\mu)_{\mu=1,\dots,N}$
determined by solving the equation
$$
2\ctil(E_\mu) +1=0   \tag{2.28}
$$
for $\{E_\mu\}_{\mu=1, \dots, N}$ and substituting in
$$
\zeta_\mu =\sqrt{-\frac{\Ptil(E_\mu)}{E_\mu \atil(E_\mu)}}=
\frac{2\btil(E_\mu)}{\sqrt{E_\mu}}.  \tag{2.29}
$$
This defines the zeroes of the eigenvectors of $\LL(\l)$ on the spectral curve
$\wt{\CC}$. In terms of these, we have:
$$
u=-i\ln\left(2\prod_{\mu=1}^N E_\mu\right) +\e\pi  \tag{2.30}
$$
where $\e=1,0$ for $N$ even or odd, respectively.
The $\{E_\mu,\zeta_\mu\}_{\mu=1,\dots,N}$ may be viewed as coordinate functions
on the coadjoint orbit through $\NN = \wh{J} \in \wh{\frak{su}}^*_R(2)$, and
the following result  is key to the linearization
of the flows (cf. {\bf [HW2]} for details):
\proclaim{\bf Proposition}
The functions $(E_\mu,\zeta_\mu)_{\mu=1,\dots, N}$ form a Darboux coordinate
system on the coadjoint orbit passing through $\Cal N(\l)$. The corresponding
orbital symplectic form is
$$
\o=\sum_{\mu=1}^{N} dE_\mu\wedge d\zeta_\mu=-d\theta. \tag{2.31}
$$
\endproclaim
This may be seen either by an explicit coordinate transformation from (2.11),
by verifying the following expression for the canonical $1$--form
$$
 \theta = 4\sum_{i=1}^p(\varphi_i d\bar\gamma_i-\gamma_i
d\bar\varphi_i)+4i\sum_{j=2p+1}^N\gamma_j d\bar\gamma_j=\sum_{\mu=1}^{N}
\zeta_\mu dE_\mu ,
\tag{2.32}
$$ or
as an application of the general ``spectral Darboux coordinates'' theorem of
{\bf [AHH3]}.

It now follows directly from the definition of the coordinates
$(E_\mu,\zeta_\mu)_{\mu=1,\dots, N}$  that on the invariant level sets
(Lagrangian manifolds) determined by  $\{P_i=c_i\}_{i=0}^{N-1}$,
the $1$--form $\theta\vert_{\{P_i =c_i\}}=dS$ may be integrated to yield the
Liouville  generating function
$$
S(P_i,E_\mu)=\sum_{\mu=1}^{N} \int_{E_0}^{E_\mu}
\sqrt{-\frac{\Ptil(E)}{E\atil(E)}}dE.
\tag{2.33}
$$
Within the new canonical coordinate system $\{Q_i, P_i\}_{i=0, \dots N-1}$
defined by
$$
Q_i=\frac{\partial S}{\partial P_i}= -\frac 12
\sum_{\mu=1}^{N} \int_{E_0}^{E_\mu}
\frac{E^i}{\sqrt{-E\atil(E)\Ptil(E)}}dE,   \tag{2.34}
$$
the flows for all the Hamiltonians in the spectral ring generated by the
$P_i$'s are then linear. In particular, integrating
Hamilton's  equations for $H_\xi,H_\eta$ gives
 $$
\sum_{\mu=1}^{N} \int_{E_0}^{E_\mu} \frac{E^i}{\sqrt{-E\atil(E)\Ptil(E)}}dE
=C_i+2\delta_{i,0}\xi-2\delta_{i,N-1}\eta. \tag{2.35}
$$
This may be interpreted as the linearizing Abel map to the Jacobi variety of
$\Ctil$.  By a standard inversion procedure (cf.~{\bf [HW2]}, the function $u$
may be explicitly expressed in terms of quotients of the associated theta
functions:
$$
u=-2i\ln\frac{\Theta(A(0)-U\eta-V\xi-B-\kappa)}
{\Theta(A(\infty)-U\eta-V\xi-B-\kappa)}
+ c ,
\tag{2.36}
$$
where the constant vectors $A,U,V,B \in \bold{C}^N$ are obtained from those
defined by the slopes and integration constants on the RHS of eq.~(2.35) by
applying the matrix that transforms the differentials appearing in the
integrands on the LHS to a normalized canonical basis of abelian differentials.

\newpage \phantom{0}
%%%%%%%%%%%%%%%%%%%%%%%%%%%%%%%%% References %%%%%%%%%%%%%%%%%%%%%%%%%%%%%%%
 \centerline{\bf References}
\bigskip {\smaller
\item{\bf [AHH1]} Adams, M.~R., Harnad, J.\  and Hurtubise, J., ``Isospectral
Hamiltonian Flows in Finite and Infinite Dimensions II.  Integration of
Flows,''
 {\it Commun.\  Math.\  Phys.\/} {\bf 134}, 555--585 (1990).
\item{\bf [AHH2]} Adams, M.~R., Harnad, J.\  and Hurtubise, J.,
``Dual Moment Maps to Loop Algebras,''  {\it Lett.\  Math.\  Phys.\/} {\bf 20},
 294--308 (1990).
\item{\bf [AHH3]}  Adams, M.~R., Harnad,~J. and  Hurtubise,~J., ``Darboux
Coordinates and Liouville-Arnold Integration  in Loop Algebras,''   preprint
CRM  (1992) (to appear in {\it Commun.\  Math.\  Phys.\/} (1993)).
\item{\bf [AHP]} Adams, M.~R., Harnad, J.\  and Previato, E., ``Isospectral
Hamiltonian Flows in Finite and Infinite Dimensions I. Generalised Moser
Systems
and Moment Maps into Loop Algebras,''   {\it Commun.\ Math.\  Phys.\/}  {\bf
117}, 451--500 (1988).
 \item{\bf[AvM]} Adler, M.\ and van Moerbeke, P., ``Completely Integrable
Systems, Euclidean Lie Algebras, and Curves,''  {\it Adv.\  Math.\/}
{\bf 38}, 267--317 (1980); ``Linearization of Hamiltonian Systems, Jacobi
Varieties and Representation Theory,''  {\it ibid.} {\bf 38}, 318--379 (1980).
\item{\bf [Du]} Dubrovin, B.A., ``Theta Functions and Nonlinear Equations'',
 {\it Russ\. Math\. Surv\.} {\bf 36}, 11-92 (1981).
\item{\bf [HW1]} Harnad, J., and Wisse, M.-A., ``Matrix Nonlinear Schr\"odinger
Equations and Moment Maps into Loop Algebras'',
 {\it J.~Math.~Phys.} {\bf 33}, 4164--4176 (1992).
\item{\bf [HW2]} Harnad, J. and Wisse, M.-A.  ``Isospectral Flow in Loop
Algebras and Quasiperiodic Solutions of the Sine-Gordon Equation'',
preprint CRM-1831 (1992), hep-th/9210077.
\item{\bf [KN]} Krichever, I.M.and  Novikov, S.P.,  ``Holomorphic Bundles over
Algebraic Curves and Nonlinear Equations'', {\it Russ\. Math.~ Surveys}
{\bf 32}, 53-79 (1980).
\item{\bf [S]} Semenov-Tian-Shansky, M.~A.,
``What is a Classical R-Matrix,''
{\it Funct.\  Anal.\  Appl.\/} {\bf 17} 259--272 (1983);
``Dressing Transformations and Poisson Group Actions,''
{\it Publ.\  RIMS Kyoto Univ.\/} {\bf 21} 1237--1260 (1985).
\item{\bf [TW]} Tafel, J., and Wisse, M.-A.,  ``Loop Algebra Approach to
Generalized Sine-Gordon Equation'',  preprint CRM (1993).
\item{\bf [W1]} Wisse, M.-A.,  ``Quasi-Periodic Solutions for
Matrix Nonlinear Schr\"odinger Equations'', \newline {\it J.~Math.~Phys.}
{\bf 33}, 3694--3699 (1992).
 \item{\bf [W2]} Wisse, M.-A.,  ``Darboux Coordinates and Isospectral
Hamiltonian Flows for the Massive Thirring Model'', preprint CRM (1993).
\medskip }
\vfill \eject
\enddocument